\def\be{\begin{equation}}
\def\ee{\end{equation}}
\def\ba{\begin{array}}
\def\ea{\end{array}}

\def\Rb{{I\!\! R}}

\def\Cb{\ \hbox{\vrule width 0.6pt height 6pt depth 0pt
		      \hskip -3.2 pt} C}
\documentstyle[12pt]{article}
\begin{document}
\parskip=4pt
\parindent=18pt
\baselineskip=22pt
\setcounter{page}{1}
\centerline{\Large\bf Separability of Rank Two Quantum States}
\vspace{6ex}
\begin{center}
Sergio Albeverio \footnote{ SFB 256; SFB 237; BiBoS; CERFIM(Locarno); Acc. Arch.; USI(Mendriso)

~~~e-mail: albeverio@uni-bonn.de},
 Shao-Ming Fei \footnote{ Institute of Applied Mathematics, Chinese Academy of Science, Beijing

~~~e-mail: fei@uni-bonn.de}, 
 Debashish Goswami \footnote{ Alexander von Humboldt Fellow

~~~e-mail: goswamid@wiener.iam.uni-bonn.de}\\
 Institut f{\"u}r Angewandte Mathematik, Universit{\"a}t Bonn, D-53115, Bonn.\\
 \end{center} 
\vskip 1 true cm
\parindent=18pt
\parskip=6pt
\begin{center}
\begin{minipage}{5in}
\vspace{3ex}
\centerline{\large Abstract}
\vspace{4ex}
Explicit sufficient  and necessary conditions for
separability of higher dimensional quantum systems
with rank two density matrices are given. 
A nonseparability inequality is also presented, for the case 
where one of the eigenvectors corresponding to nonzero eigenvalues
is a maximally entangled state.
\bigskip
\medskip
\bigskip
\medskip

\vfill

\end{minipage}
\end{center}

\newpage

Quantum entanglement has been investigated for decades
because of its importance in the foundations of quantum mechanics,
 particularly in relation with quantum nonseparability and the
violation of Bell's inequalities \cite{Bell}.  
Recently it has been viewed also as a potentially useful resource
for communication, information processing and
quantum computing \cite{DiVincenzo}, such as
for the investigation of quantum teleportation \cite{teleport,teleport1},
dense coding \cite{dense}, decoherence in quantum computers and
the evaluation of quantum cryptographic schemes \cite{crypto}.

Due to recent works
by Peres \cite{Pe96} and Horodecki et al \cite{Ho96} there exist a simple
criterion allowing one to judge, whether a given density matrix $\rho$, 
representing a $2 \times 2 $ or $2 \times 3$ composite system, is
separable. Nevertheless,
the general problem of finding sufficient and necessary conditions for
separability in higher dimensions remains open 
(see e.g. \cite{LBCKKSST,H300} and references therein).

A general condition for separability of a quantum state
could  in principle be obtained from
the measure of entanglement. To quantify the degree of entanglement
a number of entanglement measures,
such as the entanglement of formation and distillation
\cite{Bennett96a,BBPS,Vedral}, negativity \cite{Pe96,Zyczkowski98a}
and relative entropy \cite{Vedral,sw}, have been proposed
for bipartite states [6,8,15-18].
However most proposed measures of
entanglement involve extremizations which
are difficult to handle analytically.
For instance, the ``entanglement of formation'' \cite{Bennett96a}, 
which is intended to
quantify the amount of quantum communication required to create a
given state, is 
defined for arbitrary dimension, but  so far no
explicit analytic formulae for entanglement of formation
have been found for systems larger than a
pair of qubits (spin-$1\over 2$ particles).
For the case of a pair of qubits, it has been shown
that the entanglement of formation can be expressed as a 
monotonically increasing function of the ``concurrence'' variable $C$. This
function ranges from 0 to 1 as $C$ goes from 0 to 1, 
so that one can take the
concurrence as a measure of entanglement in its own right
\cite{HillWootters}.

In this letter, we introduce a generalized concurrence and
study  sufficient and necessary conditions for
separability of higher dimensional quantum systems.
In particular, we consider density matrices with rank two.
The separability condition for these kind of mixed states
in arbitrary dimensions is explicitly given.
In addition, we present a non separability inequality valid in the case where
one of the eigenvectors, corresponding to nonzero eigenvalues,
of a density matrix is a maximally
entangled state.

We first introduce a generalized concurrence $C_N$ for
$N$-dimensional quantum states. Let $H$ be an $N$-dimensional
complex Hilbert space, with
$e_i$, $i=1,...,N$, as an   orthonormal basis.
A general pure state on $H\otimes H$ is  of the form,
\begin{equation}
\vert\Psi\rangle=\sum_{i,j=1}^N a_{ij}e_i\otimes e_j,~~~~~~a_{ij}\in\Cb
\end{equation}
with the normalization $\displaystyle\sum_{i,j=1}^N a_{ij}a_{ij}^\ast=1$
($\ast$ denoting complex conjugation).

Let $U$ denote a unitary transformation 
on the Hilbert space ${\cal H}$, such that
\begin{equation}\label{unit}
U e_i \mapsto \sum_{j=1}^N b_{ij}e_j,~~~~~~b_{ij}\in\Cb
\end{equation}
and $\displaystyle\sum_{j=1}^N b_{ij}b_{kj}^\ast=\delta_{ik}$ 
(with $\delta_{ik} $ the usual Kronecker's symbol).
We call a quantity an invariant associated with the state
$\vert\Psi\rangle$ if it is invariant under all local unitary 
transformations, i.e. all maps from $H \otimes H$ to
itself  of the form 
$U\otimes U$. Let $A$ denote the matrix given by $(A)_{ij}=a_{ij}$.
By generalizing the results of analysis on invariants
for qubits \cite{Linden}, we can show that the following quantities
are invariants under local unitary transformations:
\begin{equation}\label{I}
I_\alpha=Tr(AA^\dag)^{\alpha+1},~~~~~~~~~~~\alpha=0,1,...,N-1;
\end{equation}
(with $A^\dag$ the adjoint of the matrix $A$).

We define the generalized concurrence to be:
\be\label{cn}
C_N=\sqrt{\frac{N}{N-1}(I_0^2-I_1)}=\sqrt{\displaystyle
\frac{N}{2(N-1)}\sum_{i,j,k,m=1}^N \vert
a_{ik}a_{jm}-a_{im}a_{jk}\vert^2}\,.
\ee
For $N=2$ we have $C_2=2|a_{11}a_{22}-a_{12}a_{21}\vert$, which is just
the definition of concurrence for a pure state of 
two qubits \cite{HillWootters}.

For general $N$, we see that when
the state $\vert\Psi\rangle$ is factorizable in the sense that  
$a_{ij}=a_ib_j$ for some $a_i$, $b_j\in\Cb$, $i=1,...,N$, $C_N$ is zero.
When $\vert\Psi\rangle$ is maximally entangled,
 $|a_{ii}|=1/\sqrt{N}$, $a_{ij}=0$ for $i \neq j$, $C_N$ is one.
In terms of the Schmidt decomposition, a given
$\vert\Psi\rangle$ can always be written in the form, possibly by changing the  
 orthonormal basis $\{e_i\}$, $i=1,...,N$:
$$
\vert\Psi\rangle=\sum_{i=1}^N \sqrt{\Lambda_i}e_i\otimes e_i,
$$
where $\displaystyle\sum_{i=1}^N \Lambda_i=1$, $\Lambda_i\geq 0$.
The invariants are then of the form
$$
I_\alpha=\sum_{i=1}^N \Lambda_i^{\alpha +1},~~~~~~~\alpha=0,...,N-1.
$$
We note that
$$
I_1=\sum_{i=1}^N \Lambda_i^2=I_0^2-\sum_{i \neq j}^N \Lambda_i\Lambda_j.
$$
Therefore $C_N=0$ implies that 
$\displaystyle\sum_{i \neq j}^N \Lambda_i \Lambda_j=0$.
As $\Lambda_i\geq 0$ and $\displaystyle\sum_{i=1}^N \Lambda_i=1$,
we have that in this case only 
one of the $\Lambda_i,i=1,...N$ is  equal to $1$ and all other ones  are 
zero. Hence $C_N=0$ implies that $\vert\Psi\rangle$ is separable.

If $C_N=1$, we have $\displaystyle\sum_{i \neq j}^N \Lambda_i
\Lambda_j=\frac{N-1}{N}$,
which is equivalent to the condition
$\displaystyle\sum_{i=1}^N \Lambda_i^2=1/N$, according to the normalization
$\displaystyle\sum_{i=1}^N \Lambda_i=1$.
The equation $\displaystyle\sum_{i=1}^N \Lambda_i^2=1/N$ describes 
an $(N-1)$-dimensional
sphere in $\Rb^N$ with radius $1/\sqrt{N}$, 
whereas $\displaystyle\sum_{i=1}^N \Lambda_i=1$
is a hyperplane in $\Rb^N$. These hypersurfaces  
have only one contact point at 
$\Lambda_i=1/N$, $i=1,...,N$. Therefore $C_N=1$
implies that $\vert\Psi\rangle$ is maximally entangled.
We remark that the above properties of $C_N$ do not mean that
$C_N$ is in general a suitable measure for general $N$-dimensional
bipartite quantum pure states. It can however be shown that when the
matrix $AA^\dag$ has only two different nonzero eigenvalues, the
entanglement of formation is a monotonically
increasing function of $C_N$, thus  $C_N$ can indeed be interpreted as 
a measure of entanglement in this case.

Let $\rho$ be a rank two state in $H\otimes H$, with
$|E_1\rangle$, $|E_2\rangle$ being its two orthonormal eigenvectors
corresponding to the two nonzero eigenvalues: 
\be\label{rho}
\rho=p|E_1\rangle\langle E_1| +q|E_2\rangle\langle E_2|,
\ee
where $q=1-p\in (0,1)$.
Generally $|E_k\rangle=\displaystyle\sum_{i,j=1}^N a_{ij}^k e_i\otimes e_j$,
$a_{ij}^k\in\Cb$,
with normalization $\displaystyle\sum_{i,j=1}^N a_{ij}^k (a_{ij}^{k})^\ast=1$, $k=1,2$.
We would like to give an explicit 
algebraic condition for the separability of the above state in terms of 
$a^k_{ij}, k=1,2$.
   
With the notations:
$$
\ba{l}
\alpha_{ij}^{kl}=a^2_{ij}a^2_{kl}-a^2_{il}a^2_{kj},~~~
\gamma_{ij}^{kl}=a^1_{ij}a^1_{kl}-a^1_{il}a^1_{kj}\\[3mm]
\beta_{ij}^{kl}=a^1_{ij}a^2_{kl}+a^2_{ij}a^1_{kl}-a^2_{il}a^1_{kj}
-a^1_{il}a^2_{kj},
\ea
$$
we have the following conclusion:
	 
\noindent {\bf [Theorem 1].} 
If all $a^k_{ij}$ are real, $\rho$ is separable if and only if
 {\bf one} of the following  quantities ($\Delta_1$
or $\Delta_2)$ is zero :
$$
\Delta_1=\sum_{ijkl} | \gamma_{ij}^{kl}-(1-p^{-1})\alpha_{ij}^{kl}|^2
+\sum_{ijklmn} | \beta_{ij}^{kl} \alpha_{mn}^{kl}-\alpha_{ij}^{kl}
\beta_{mn}^{kl}|^2;
$$
$$ 
\Delta_2=\sum_{ijkl} | \gamma_{ij}^{kl}+(1-p^{-1})\alpha_{ij}^{kl}|^2+
\sum_{ijkl} |\beta_{ij}^{kl}|^2;
$$
or equivalently one of the following two sets of relations ((6) or (7)) hold:
\be\label{t11}
\gamma_{ij}^{kl}=(1-p^{-1})\alpha_{ij}^{kl}\,,~~
\beta_{ij}^{kl}
\alpha_{mn}^{kl}=\alpha_{ij}^{kl}\beta_{mn}^{kl}\,,~~
\forall\, i,j,k,l,m,n.
\ee
\be\label{t12}
\gamma_{ij}^{kl}=-(1-p^{-1})\alpha_{ij}^{kl}\,,
~~\beta_{ij}^{kl}=0\,,~~\forall\, i,j,k,l.
\ee

{\sf [Proof].} From the discussions above we have that
a state $\vert\Psi\rangle=\displaystyle\sum_{i,j=1}^N a_ {ij}e_i\otimes e_j$ 
is separable if and only if the generalized concurrence $C_N$ is zero,
i.e., 
\be\label{eq1}
a_{ij}a_{kl}=a_{il}a_{kj}, ~~~\forall\, i,j,k,l.
\ee
Therefore a vector of the 
form $\vert E_1\rangle +\lambda \vert E_2\rangle, \lambda \in \Cb$ is 
separable if and only if $\lambda$ is a common root of the following
equation set $Eq_{ij;kl}$:
\be\label{eq2}
\alpha_{ij}^{kl} \lambda^2
+\beta_{ij}^{kl} \lambda+\gamma_{ij}^{kl}=0, ~~~\forall\, i,j,k,l.
\ee
  
We first prove the necessity part of the theorem.
Suppose that 
$\rho=\displaystyle\sum_{t=1}^l p^\prime_t |U_t\rangle\langle U_t|,$ 
with $l$ some positive integer
and $0 <p^\prime_t <1$, $\sum p^\prime_t =1$, gives a decomposition
of $\rho $ in terms of separable normalized vectors 
$U_t$ (where $|U_t\rangle\langle U_t|$ means the projection onto the
vector $U_t$). As each of these 
vectors must be in the range of $\rho$ (since for each fixed $t$,
$|U_t\rangle\langle U_t| \leq {p_t^\prime}^{-1}\rho$), we can write them
as linear combinations 
of the two eigenvectors $\vert E_1\rangle$ and $\vert E_2\rangle$ 
which span the range of $\rho$. So let us assume that $U_t=c_1^t \vert
E_1\rangle +c_2^t \vert E_2\rangle$ (for some $c_1^t,c^t_2 \in \Cb$). 
We deal with the problem in two cases:

{\raggedright Case 1. $\vert E_2\rangle$ is not separable}

In this case, since $c_1^t$ can not be $0$,  $E_1+\lambda_t E_2$ 
will be a separable vector, where $\lambda_t=c_2^t/c_1^t$.
Clearly, not all $\lambda_t$ s can be equal, otherwise 
all the $U_t$ s would be constant multiples of a fixed vector, and
$\rho$ would be of rank 1. Hence there must be at least 
two distinct choices of $\lambda_t$. On the other hand, 
as $\vert E_2\rangle$ is not separable, its generalized concurrence $C_N$ is
not zero. Hence there is some $i_0, j_0, k_0, l_0$
such that $\alpha_{i_0j_0}^{k_0l_0}\neq 0$, i.e. the relation 
$Eq_{i_0j_0;k_0l_0}$ is indeed a quadratic equation.
It must have exactly two roots, say $\lambda^{(1)},
\lambda^{(2)}$, and so these two values are the only possible choices
for the $\lambda_t$'s.
But in order that there is not only one possible choice, the above two
roots must
be different. And all the relations $Eq_{ij;kl}$ have these different
two roots. Consider for any $i,j,k,l$,

\noindent Situation 1: If $\alpha_{ij}^{kl}\neq 0$, 
the corresponding relation (\ref{eq2}) 
is then not an identity. All the quadratic equations
in the equation set $Eq_{ij;kl}$ have
the same two distinct roots. In other words,
$Eq_{ij;kl}$ and $Eq_{i_0j_0;k_0l_0}$ have the same roots.
From the standard theory of quadratic equations, we have
\be\label{i}
\beta_{ij}^{kl}\alpha_{i_0j_0}^{k_0l_0}=\beta_{i_0j_0}^{k_0l_0}
 \alpha_{ij}^{kl},
\ee
\be\label{ii}
\gamma_{ij}^{kl}\alpha_{i_0j_0}^{k_0l_0}=\gamma_{i_0j_0}^{k_0l_0}
\alpha_{ij}^{kl}.
\ee

\noindent Situation 2 : If $\alpha_{ij}^{kl}=0$, then the equations
 $Eq_{ij;kl}$ become identities, i.e. $\beta_{ij}^{kl}$ and 
$\gamma_{ij}^{kl} $ must be $0$ too, because otherwise
at least one of the relations $Eq_{ij;kl}$ would be a linear equation, 
and so could not be satisfied by 
two distinct roots. Thus in this case (\ref{i}) and 
(\ref{ii}) also hold (as  both  sides in both equations vanish). 
				   
Thus we have obtained (\ref{i}) and 
(\ref{ii}) for arbitrary $ijkl$ and for
all $i_0j_0k_0l_0$ such that $\alpha_{i_0j_0}^{k_0l_0}
\neq 0$. As (\ref{i}) and (\ref{ii}) are trivially valid 
whenever $ijkl$ and $i_0j_0k_0l_0$ are such that $\alpha
_{ij}^{kl}=\alpha_{i_0j_0}^{k_0l_0}=0$, we have indeed established
(\ref{i}) and (\ref{ii}) for all quadruplets $ijkl$, $i_0j_0k_0l_0$.                                          
 
We now denote the above two distinct roots, which are common
to all of the equations $Eq_{ij;kl}$,
by $\mu_1,\mu_2$, with the convention that in case the roots are real, 
$\mu_2$ will denote the larger one. Each vector $U_t$ is either of the
form $\tilde{E_1}=\frac{E_1+\mu_1E_2}{\sqrt{1+|\mu_1|^2}}$ or of the
form $\tilde{E_2}=\frac{E_1+\mu_2E_2}{\sqrt{1+|\mu_2|^2}} $.
Therefore we can write $\rho$ 
as, $\rho=p^\prime |\tilde{E_1}\rangle\langle\tilde{E_2}|
+(1-p^\prime)|\tilde{E_2}\rangle
\langle\tilde{E_2}|$, with $0 < p^\prime < 1.$  
Comparing with the 
coefficients of $|E_k\rangle\langle E_l|$, $k,l=1,2$ in the
expression (\ref{rho}), we get that the above 
decomposition of $\rho$ is equivalent to the following two relations:
\be\label{ppeq}
\frac{p^\prime}{1+|\mu_1|^2}+\frac{1-p^\prime}{1+|\mu_2|^2}=p,~~~
\frac{\mu_1 p^\prime}{1+|\mu_1|^2}+\frac{\mu_2(1-
p^\prime)}{1+|\mu_2|^2}=0.
\ee
Solving (\ref{ppeq}) for $p$ and $p^\prime$ we get
\be\label{pp}
p=(1-\mu_1 \mu_2 \frac{\bar{z}}{z})^{-1},~~~
p^\prime=\frac{\mu_2(1+|\mu_1|^2)}
{z-\mu_1 \mu_2 \bar{z}},
\ee
where $z=\mu_2-\mu_1$.
					 
As $a^k_{ij}$s are real numbers,
$\mu_1,\mu_2$ are roots of a quadratic equation with real coefficients. 
For a quadratic equation $a x^2 + b x + c =0$ with
$a,b,c\in\Rb$, $a\neq 0$, and roots $\alpha,\beta$ with $\alpha \neq \beta$, we have that 
$\frac{\bar{\alpha}-\bar{\beta}}{\alpha-\beta}$ is either $+1$ or $-1$, 
depending on whether $b^2-4ac$ is positive or negative respectively. 
Hence in our case we have that
$\mu_1 \mu_2=+(1-p^{-1})$ or $-(1-p^{-1}).$ 
On the other hand, since $\mu_1\mu_2$
is real, the solution for $p^\prime$ in (\ref{pp}) implies 
that $\frac{\mu_2}{\mu_2-\mu_1} $ is real, which is possible if and 
only if either the roots are both  real or the roots are
both purely imaginary.

For the  the case where  the roots are both purely imaginary, we have,
$\mu_1 \mu_2=-(1-p^{-1})$; and a direct simplification from (\ref{pp})
gives $p^\prime=\frac{1}{2}$.
The condition for having purely imaginary roots of quadratic equations
gives that $\beta^{kl}_{ij}=0$, $\forall ijkl$. 

In the case where the roots are real, $\mu_2$ being the larger root by assumption,
we have $\mu_1\mu_2=(1-p^{-1})$. By replacing $|\mu_1|^2$ by $\mu_1^2$
in the expression of $p^\prime$ in (\ref{pp}), we get
that the condition that $p^\prime$ takes values between $0$ and $1$ is 
equivalent to $\mu_2>0$, $\mu_1<0$, which is trivially valid since
$\mu_2>\mu_1$ and $\mu_1\mu_2 <0$ in this case. 

It remains to observe case that $\mu_1 \mu_2$ is nothing but the ratio
$\frac{\gamma_{i_0j_0}^{k_0l_0}}{\alpha_{i_0j_0}^{k_0l_0}}$ which is
either $1-p^{-1}$ or $-(1-p^{-1})$.
Taking into account  relation (\ref{ii}), one concludes that
either $\gamma_{ij}^{kl}=(1-p^{-1}) \alpha_{ij}^{kl}$ or 
$\gamma_{ij}^{kl}=-(1-p^{-1}) \alpha_{ij}^{kl}$ for all $i,j,k,l$.
It then follows that in the second case $\beta_{ij}^{kl}=0$ for 
all $i,j,k,l$. This completes
the proof for the necessity part in the case 1.

{\raggedright Case 2. $\vert E_2\rangle$ is separable}
							 
In this case from (\ref{eq1}) we have
$\alpha_{ij}^{kl} =0$, $\forall i,j,k,l$. 
Since not all of the $U_t$'s can be  multiples  of
$\vert E_2\rangle$, we must have at least one choice of $\lambda$ such that
$E_1+\lambda E_2$ is separable. This $\lambda$ must be a common 
root to all equations $Eq_{ij;kl}$ as before.
All these equations are now linear ones.
Excluding the trivial case that all of them are identically $0$ 
(which means that $\vert E_1\rangle$ is also separable), we see that
there is only one possible choice of $\lambda$ (which is the solution to
a  nontrivial linear equation).
Then $\rho$ can be expressed as $\rho=p^{\prime \prime}
|E_2\rangle\langle E_2|+(1-p^{\prime \prime} )\frac{|E_1+\lambda E_2\rangle
\langle E_1+\lambda
E_2|}{1+|\lambda|^2}$. That is $p^{\prime \prime}=1$, 
which is clearly a contradiction. Thus in case where $\vert E_2\rangle$ is
separable, we must have $\vert E_1\rangle$ to be separable too 
 in order for  $\rho$ to be  separable. It is clear that
in this case the conditions of the theorem hold. 

Now we prove the sufficiency part for the theorem.
If either (\ref{t11}) or (\ref{t12}) holds, from the previous
analysis it is clear that the equations $Eq_{ij;kl}$ have common roots.
If $\vert E_2\rangle$ is not separable,
then not all of these equations are identities. And there are at most
two common roots. Furthermore, if (\ref{t11}) holds, 
the product of the two roots must be $(1-p^{-1})<0$,
so that the two roots be real and unequal, with opposite signs. Denoting
them by $\mu_1$, $\mu_2$ as before, we can easily get 
from the arguments given earlier an explicit decomposition of 
$\rho$ in terms of separable pure states. Similarly,
if (\ref{t12}) holds, the two roots must be purely 
imaginary (since $\beta_{ij}^{kl}=0$) and the 
rest also follows easily from our earlier analysis. 
Finally, if $\vert E_2\rangle$ is separable, either 
of the conditions (\ref{t11}) or (\ref{t12}) 
forces $\gamma_{ij}^{kl}$ to be $0$ for all $ijkl$. Hence
$\vert E_1\rangle$ is separable too and $\rho$ is trivially separable.
This completes the proof.

The conclusions in Theorem 1 can easily be generalized to the 
complex case, $a_{ij}^k\in \Cb$.

\noindent {\bf [Theorem 2]}.
$\rho$ is separable if and only if there is $\theta\in\Rb$ such that  
\be\label{t21}
\gamma_{ij}^{kl}=e^{i\theta}(1-p^{-^1})\alpha_{ij}^{kl},
\ee
\be\label{t22}
\beta_{ij}^{kl}\alpha_{mn}^{kl}=\alpha_{ij}^{kl}\beta_{mn}^{kl}\,,~~~
\forall\, i,j,k,l,m,n;
\ee
and  
\be\label{t23}
\frac{\mu_2(1+|\mu_1|^2)}{z-\mu_1 \mu_2 \bar{z}}\in [0,1],
\ee
where $z=e^{i \theta}\bar{z}$, $z=\mu_2-\mu_1\neq 0$,
$\mu_1$ and $\mu_2$ are the roots of the equation $\alpha^{kl}_{ij}
\lambda^2+\beta^{kl}_{ij}\lambda+\gamma^{kl}_{ij}=0$, for some 
$i,j,k,l$ such that $\alpha^{kl}_{ij} \neq 0$.

The proof for the necessity part of the theorem is similar
to the proof of the corresponding part in Theorem 1. One only needs to note that
since $z/\bar{z}$ is of modulus $1$,
and of the form $e^{i \theta}$ for some $\theta$. From the
relations (\ref{pp}), we obtain then that 
$\frac{\mu_2(1+|\mu_1|^2)}{z-\mu_1 \mu_2 \bar{z}}$ is a real number
between $0$ and $1$.
	 
The proof of the sufficiency part for the theorem
essentially follows by observing that the arguments given in
the necessity part can be reversed. From the previous analysis,
it follows from (\ref{t21}) and (\ref{t22}) 
that the equations $Eq_{ij;kl}$ have common roots. In case all
$\alpha^{kl}_{ij}$ are $0$, by the condition 
$\gamma^{kl}_{ij}=e^{i \theta}(1-p^{-1})\alpha^{kl}_{ij}$,
we see that all $\gamma^{kl}_{ij}$ are $0$ too.
Hence both $\vert E_1\rangle$ and $\vert E_2\rangle$ are separable, and so is $\rho$.
In the other situations where some of the  $\alpha^{kl}_{ij}$'s are nonzero, 
the corresponding equations $Eq_{ij;kl}$
have exactly two roots which are 
different by the condition (\ref{t23}). Moreover,
(\ref{t23}) ensures that the explicit solution for 
the choice of $p^\prime$,
which gives the decomposition of $\rho$ in terms 
of separable states, is indeed a valid choice 
between $0$ and $1$. This completes the proof of theorem 2.

\noindent {\bf [Corollary]}.
Let $\vert E_2\rangle$ be the maximally entangled 
vector given by  $\vert E_2\rangle=\frac{1}{\sqrt{N}}
\displaystyle\sum_{i=1}^N e_i \otimes e_i$. 
For any vector $\vert E_1\rangle$ which is orthogonal to $\vert E_2\rangle$,
$\rho=p|E_1\rangle\langle E_1|+(1-p)|E_2\rangle\langle E_2|$ is not
separable for $0 <p<\frac{1}{2}$.

\noindent {\sf [Proof]}.
From the necessary condition for separability,
$\gamma^{kl}_{ij}=e^{i \theta}(1-p^{-1})\alpha^{kl}_{ij}$, we get 
that $C_{(1)}=\frac{1-p}{p}C_{(2)}$, where $C_{(1)}=
\sqrt{\frac{N}{2(N-1)}\displaystyle\sum_{i,j,k,m=1}^N
\vert\gamma^{kl}_{ij}\vert^2}$ and
$C_{(2)}=\sqrt{\frac{N}{2(N-1)}\displaystyle\sum_{i,j,k,m=1}^N
\vert\alpha^{kl}_{ij}\vert^2}$ are the generalized concurrences
associated with the states $|E_1\rangle$ and $|E_2\rangle$
respectively. As a pure 
state is separable if and only if the corresponding
generalized concurrence is zero,
a necessary condition for
separability is that $C_{(1)}$ and $C_{(2)}$ should be inversely proportional 
to the contribution of the corresponding pure state to $\rho$, i.e. the
eigenvalue corresponding to that eigenstate. From
$\frac{C_{(1)}}{C_{(2)}}=\frac{1-p}{p}\leq 1$, we have $p\geq \frac{1}{2}$.

We have studied the sufficient and necessary conditions for
separability of higher dimensional quantum systems
of density matrices with rank two, in terms of the generalized concurrence.
Our approach concerns the eigenstates of a given density matrix $\rho$
(thus results as in our corollary are straightforward to deduce).
It gives a direct minimal decomposition of  
separable states in terms of pure product states. The method can also be used
to check whether it is possible to 
 construct a separable mixed state from two given pure states.

Another elegant approach, based on the positive partial transpose (PPT) condition,
dealing with the sufficient and necessary criteria for separability of 
higher dimensional quantum systems was presented in \cite{horod} \footnote{We would
like to thank the referee for introducing  this paper to us.}. For a given density 
matrix $\rho$, one can judge its separabilty by calculating its 
supporting dimensions ($\rho$ is supported on $M\times N$ if this is the 
smallest product Hilbert space on which $\rho$ can act), checking the
positiveness of partial transposed $\rho$ and comparing the rank of
$\rho$ with dimensions $M,\,N$.

\vspace{2.5ex}
{\bf Acknowledgement:}\\
One of the authors (D. Goswami) would like to thank Prof. K. R. Parthasarathy
 from I.S.I. (New Delhi, India) for his motivating lecture-series on quantum
 codes, quantum computation and many related topics delivered 
  in the academic year 1999-2000. He would also express his gratefulness to 
 Prof. S. Albeverio for inviting him to I.A.M. (Bonn) and the Av. Humboldt Foundation for 
 the research fellowship.

\end{document}